\documentclass[natbib]{emulateapj}
\input epsf

\def\s2n{S^{\prime}/N}

\begin{document}
\title{High Resolution Mapping of Interstellar Clouds by Near--IR Scattering}
\author{Paolo Padoan\altaffilmark{1}, Mika Juvela\altaffilmark{2},
Veli-Matti Pelkonen\altaffilmark{2}}
\altaffiltext{1}{Department of Physics, University of California, San Diego, 
CASS/UCSD 0424, 9500 Gilman Drive, La Jolla, CA 92093-0424; ppadoan@ucsd.edu}
\altaffiltext{2}{Helsinki University Observatory, FIN-00014, University of Helsinki, Finland}

\begin{abstract}

We discuss the possibility of mapping interstellar clouds at unprecedentedly 
high spatial resolution by means of near--IR imaging of their scattered 
light. We calculate the scattering of the interstellar radiation field
by a cloud model obtained from the simulation of a supersonic turbulent
flow. Synthetic maps of scattered light are computed in the J, H and K bands 
and are found to allow an accurate estimate of column density, 
in the range of visual extinction between 1 and 20 magnitudes. We
provide a formalism to convert the intensity of scattered light at
these near--IR bands into a total gas column density. We also show
that this new method of mapping interstellar clouds is within the
capability of existing near--IR facilities, which can achieve a
spatial resolution of up to $\sim 0.1$~arcsec. This opens new
perspectives in the study of interstellar dust and gas structure on
very small scales.  The validity of the method has been recently
demonstrated by the extraordinary images of the Perseus region
obtained by \cite{Foster+05}, which motivated this investigation.

\end{abstract}

\keywords{
ISM: clouds ---  ISM: structure ---   scattering --- infrared: ISM ---  radiative transfer --- 
}

\section{Introduction}

The spatial structure of interstellar clouds is the result of the
interaction of supersonic turbulent flows with gravitational and
magnetic forces, which eventually leads to the formation of
stars. Investigations of the cloud structure can therefore provide
constraints for models of star formation, as well as indirect evidence
of the relative importance of turbulence, self--gravity and magnetic
fields.

Extended maps of interstellar clouds are usually obtained from i) the
integrated intensity of emission lines of various molecular or atomic
tracers, especially CO and HI, ii) the thermal emission of dust grains
at far--IR and sub-mm wavelengths, and iii) the extinction of the
near--IR light from background stars. Shortcomings of these methods
are well known. A given molecular line, for example, is most sensitive
to a limited range of gas density, above the critical excitation
density, and below a density where saturation sets in due to the large 
optical depth. Furthermore, molecular
abundances depend on complicated chemical networks affected by
turbulent transport and depletion on dust grains. Thermal dust
emission at far--IR and sub-mm wavelengths provides a more
straightforward picture of column density, if the dust to gas ratio is
constant and if the dust temperature can be estimated. However, both
dust temperature and optical properties of dust grains may have
significant spatial variations, due to grain growth by coagulation and
ice mantle deposition. Finally, reliable near--IR extinction maps can
be computed only for regions at intermediate Galactic latitudes and
the uncertainty on the reddening is rather large as the spectral type
of most background stars is usually unknown.

In this Letter we discuss a new method of mapping interstellar clouds,
based on the near--IR scattering of the general interstellar radiation
field. This new method avoids many shortcomings of the other methods mentioned
above. More importantly, it provides images of interstellar clouds at
an unprecedented spatial resolution, two orders of magnitude better
than any single--dish line or continuum surveys. Near--IR scattering
can be used to infer column densities within the approximate range of
$A_{\rm V}=1$ to 20~mag. Even lower column density can be probed with
shorter wavelengths, using visual and UV scattering. This study was
motivated by the extraordinary images of the Perseus region obtained
by \cite{Foster+05}.

Because of the very high spatial resolution (up to 
$\sim 0.1$~arcsec with adaptive optics), near--IR scattering images offer
exciting new perspectives in the investigation of the very small scale
structure of interstellar clouds.

\section{Turbulent Cloud Model}

\begin{figure*}[ht]
\centerline{
\includegraphics[width=8cm]{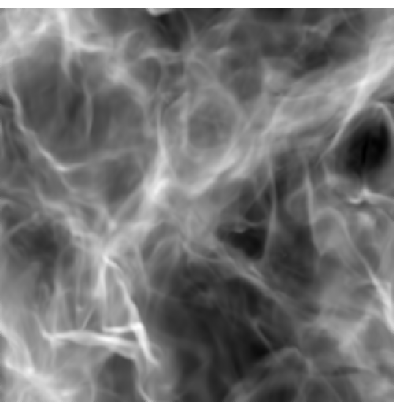}
\includegraphics[width=8cm]{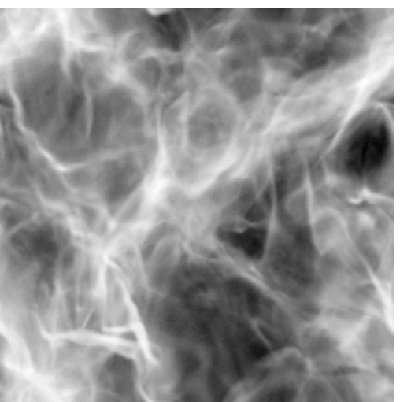}
}
\caption[]{{\em Left:} Projected density field from the super--Alfv\'enic 
MHD turbulence simulation with $M_{\rm S}=10$. {\em Right} panel: Scattered light
in the central wavelength of the H band computed for the density field
shown in the left panel.}
\label{fig1}
\end{figure*}

The density distribution of the model is the result of a three
dimensional simulation of super--Alfv\'{e}nic, compressible,
magneto--hydrodynamic (MHD) turbulence, with rms sonic Mach number of
the flow, $M_{\rm S}=10$.  The simulation is carried out on a
staggered grid of 250$^{3}$ computational cells, with periodic
boundary conditions. Turbulence is set up as an initial large scale
random and solenoidal velocity field (generated in Fourier space with
power only in the range of wavenumbers $1\le K\le 2$) and maintained
with an external large scale random and solenoidal force, correlated
at the largest scale turn--over time. The initial density and magnetic
fields are uniform and the gas is assumed to be isothermal. Details
about the numerical method are given in \cite{Padoan+Nordlund99mhd}.
 
The experiment is run for approximately 10 dynamical times in order to
achieve a statistically relaxed state. The cloud model used in this
work corresponds to the final snapshot of the simulation. The initial
rms Alfv\'{e}nic Mach number is $M_{\rm A}=10$. The volume--averaged
magnetic field strength is constant in time because of the imposed
flux conservation. The magnetic energy is instead amplified. The
initial value of the ratio of average magnetic and dynamic pressures
is $\langle P_{\rm m} \rangle _{\rm in} / \langle P_{\rm d} \rangle
_{\rm in}=0.005$, so the run is initially super--Alfv\'{e}nic.  The
value of the same ratio at later times is larger, due to the magnetic
energy amplification, but still significantly lower than unity,
$\langle P_{\rm m} \rangle _{\rm in} / \langle P_{\rm d} \rangle _{\rm
in}=0.12$.  The turbulence is therefore super--Alfv\'{e}nic at all
times.

Supersonic and super--Alfv\'{e}nic turbulence of an isothermal gas
generates a density distribution with a very strong contrast of
several orders of magnitude. It has been shown to provide a good
description of the dynamics of molecular clouds and of their highly
fragmented nature \citep[e.g.][]{Padoan+99,Padoan+01,Padoan+04}.

\section{Synthetic Maps of Near--IR Scattering}
 
To compute the radiative transfer the data cube of the density distribution is scaled
to physical units by fixing the length of the computational grid to $L=1.3$~pc and
the mean density to $\langle n _{\rm H}\rangle=1000$~cm$^{-3}$.
However, the radiative transfer calculations apply to all values of $L$ and $\langle
n _{\rm H}\rangle$ satisfying the condition that their product is constant,
$L\,\langle n _{\rm H}\rangle=1300$~pc~cm$^{-3}$.  When results are compared with
observations, also the angular size of the models (or their distance) must be fixed.

The model cloud is illuminated by an isotropic background radiation, with intensities
calculated according to  
\citet{Mathis83} model of the interstellar radiation field (ISRF). Dust
properties are taken from \citet{Draine2003a}, and correspond to
normal Milky Way dust with $R_{\rm V}=3.1$. We use the tabulated
scattering phase functions that are available on the web\footnote{{\tt
http://www.astro.princeton.edu/\~draine/dust/scat.html}}.

The flux of scattered radiation is calculated with a Monte Carlo program
\citep{Juvela2003a, Juvela2005a}, where sampling of scattered radiation is
further improved with the `peel-off' method \citep{Yusef94}. During each run,
photons of one wavelength are simulated and scattered intensity, including
multiple scatterings, is registered toward one direction. The outcoming
intensity is calculated separately for the central wavelengths of the J, H, and K
bands (1.25, 1.65, and 2.22\,$\mu$m) and for three directions perpendicular to the
faces of the cubic model cloud. The result consists of maps of scattered 
intensity, where the pixel size corresponds to the cell size of the MHD simulation. 
The number of simulated photon packages is $4\times 10^8$ per wavelength.

\section{Results}

\begin{figure*}[ht]
\includegraphics[width=\columnwidth]{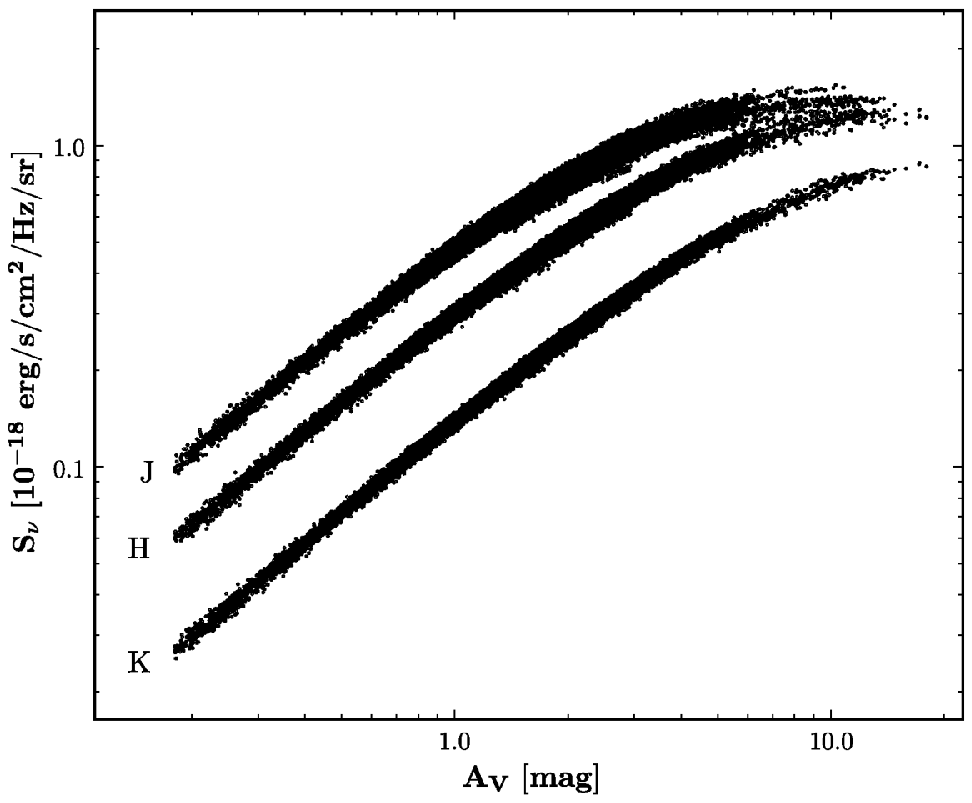}
\includegraphics[width=\columnwidth]{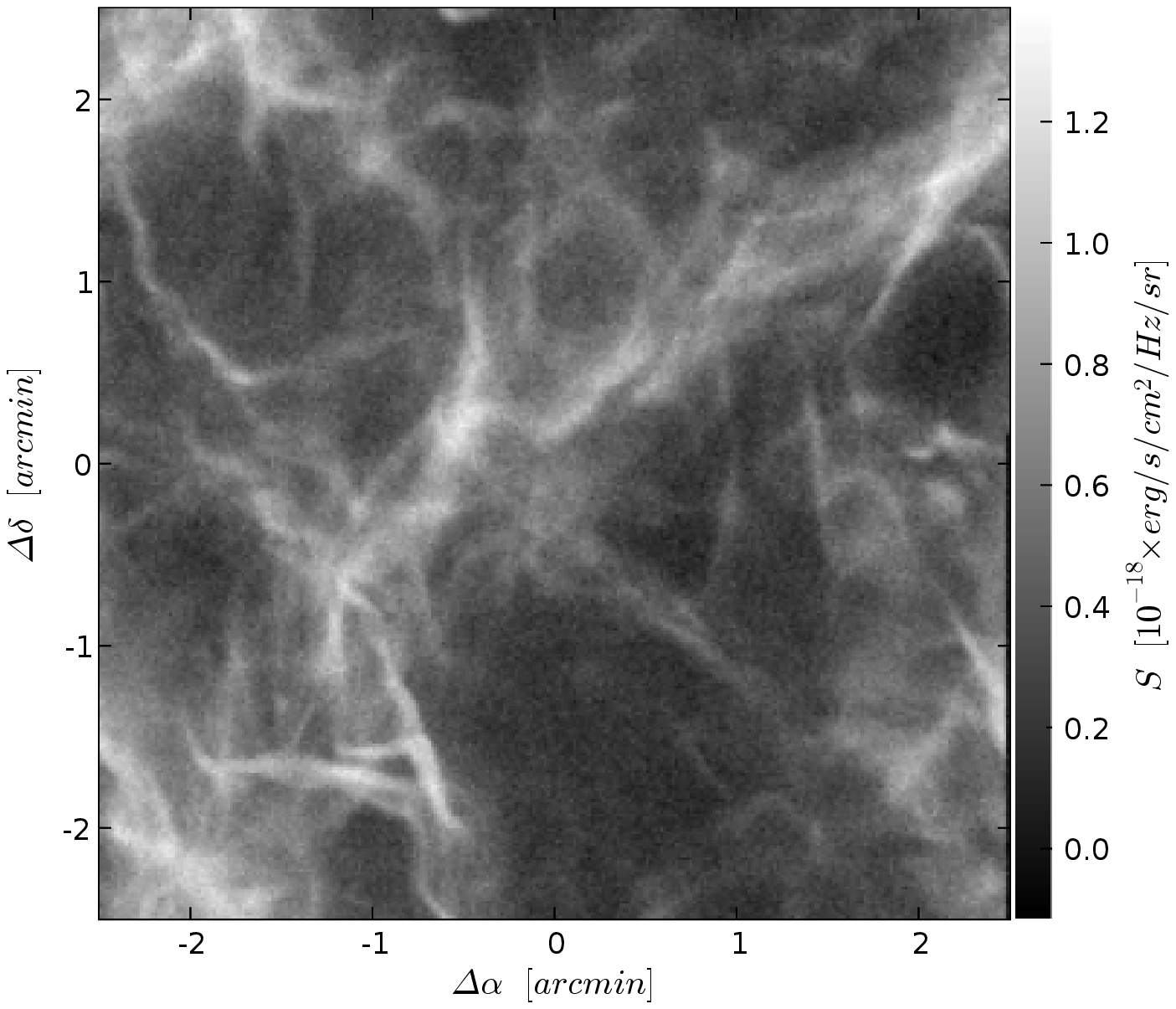}
\caption[]{
Results from numerical modeling of light scattering.
{\em Left:} J-, H-, and K-band surface brightness values as a 
function of $A_{\rm V}$. No noise was added to these values.
{\em Right:} Surface brightness of scattered light in the
H band. Noise was added to bring the signal--to--noise ratio down 
to 10 (see text).}
\label{fig2}
\end{figure*}

The left panel of Figure~\ref{fig1} shows a projection of the density field from the
turbulence simulation described above. The corresponding image of scattered light in
the H band is shown on the right panel of the same figure, for the model with mean
gas density $\langle n _{\rm H}\rangle=1000$~cm$^{-3}$. The H band scattering
reproduces well the spatial structure of the cloud model. The same is true for the J
and K bands as well (not shown for lack of space).  The scattered light tends to
saturate only in the densest regions, at approximately 5, 10 and 20  magnitudes of
visual extinction for the J, H and K bands respectively, as illustrated in the left
panel of Figure~\ref{fig2}.

Based on the radiative transfer along a single line of sight we may
assume that the observed intensity $I$ can be approximated with the function
\begin{equation}
I = a\, (1 - {\rm e}^{-bN}),
\label{eq:c1}
\end{equation}
where $N$ is the column density. The coefficients $a$ and $b$ are positive
constants defined for each band. They are related to dust properties and to the 
strength of the radiation field illuminating the cloud. To first order in $N$
(small column densities), Eq.~\ref{eq:c1} gives
\begin{equation}
I = a\,b\,N,
\label{eq:c2}
\end{equation}
so the intensity is proportional to the column density. At large values of 
column density the relation is not linear, but can still be derived by 
comparing different bands. For example, by writing Eq.~\ref{eq:c1} for the H and
K bands, $N$ can be eliminated and the H band can be expressed as a function
of the K band,
\begin{equation}
I_{\rm H} = a_{\rm H} \times (1 - (1 - \frac {I_{\rm K}}{a_{\rm K}})^{\frac 
{b_{\rm H}}{b_{\rm K}}}).
\label{eq:c3}
\end{equation}
The coefficients $a_{\rm J}$, $a_{\rm H}$, and $a_{\rm K}$ and the ratios
$b_{\rm J}/b_{\rm K}$ and $b_{\rm H}/b_{\rm K}$ can be determined by fitting
this curve to the observations (this is actually done in the
three--dimensional (J,H,K) space). 

The $b$ coefficients depend mainly on the properties of the dust grains. They
should be well defined, since dust properties are believed to be very constant
in the NIR. Their absolute values can be obtained from previous observations
of similar objects or by using stellar extinction measurements to fix the
$A_{\rm V}$ scale in the current observations. The $a$ coefficients depend on
the radiation field that illuminates the cloud. However, the spectrum of the
incoming radiation can be directly estimated using observations at low
extinction, and the field strength appears basically as a multiplicative
factor in the estimated column densities. 

The same deep observations carried out to detect the NIR scattering also 
provide color excesses for a large number of background stars. The intensity 
of the incoming radiation can be determined by comparing the stellar extinction 
with the amount of scattered light. All the coefficients can also be obtained 
from radiative transfer models based on the assumed dust properties and radiation 
field strength. Once the coefficients are determined, Eq.~\ref{eq:c1} defines, 
as a function of column density, a fixed curve in the three-dimensional (J,H,K)
space. Column densities are estimated by finding the closest point on this curve 
to each observed (J,H,K) triplet and solving for $N$ in Eq.~\ref{eq:c1}.

We have applied this method to the numerical cloud model described above.
Noise is first generated so the signal--to--noise ratios for the mean
intensities in the J, H,and K bands are 15, 10 and 7, respectively. 
The radiative transfer model gives $a_{\rm J,H,K}$ = 1.04, 0.83, 0.54$\,\times
10^{-18}$~${\rm erg\,cm^{-2}\,sr^{-1}\,Hz^{-1}}$ and $b_{\rm J,H,K}$ = 0.019,
0.012, 0.0074$\,\times 10^{-20}$~${\rm cm^2}$, based on the true column
densities and the scattered light observed toward one direction of the 
cloud model. With these coefficients, the column density for a different, 
orthogonal viewing direction is then estimated using the 'observed' surface 
brightness in that direction.
\begin{figure*}[ht]
\includegraphics[width=\columnwidth]{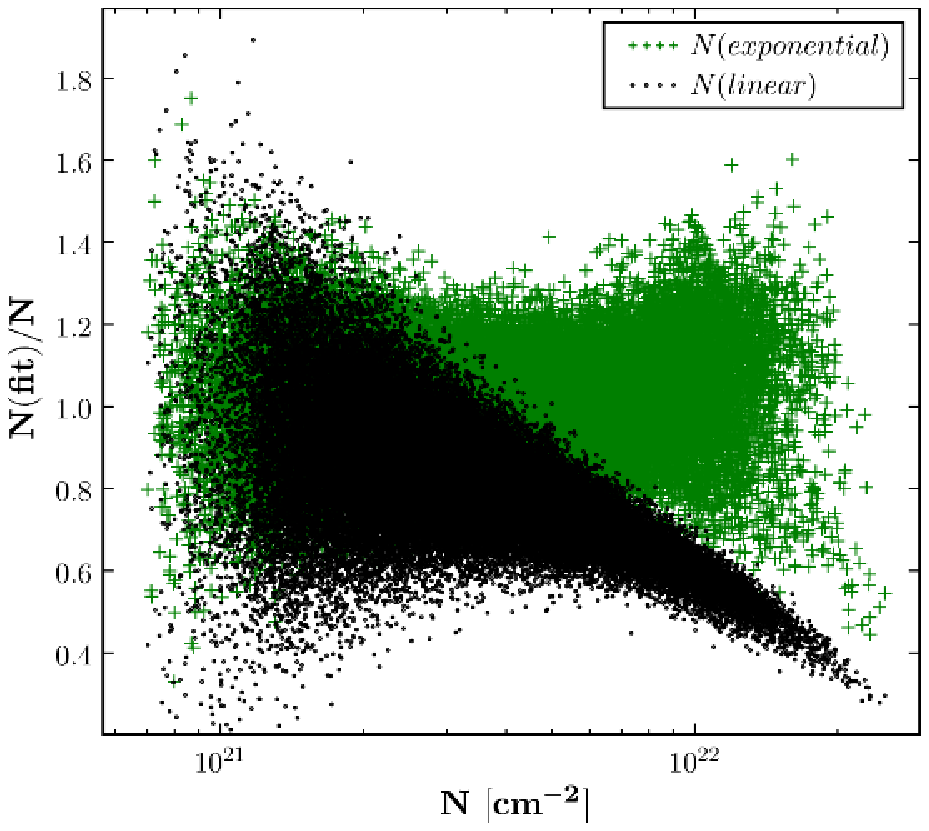}
\includegraphics[width=\columnwidth]{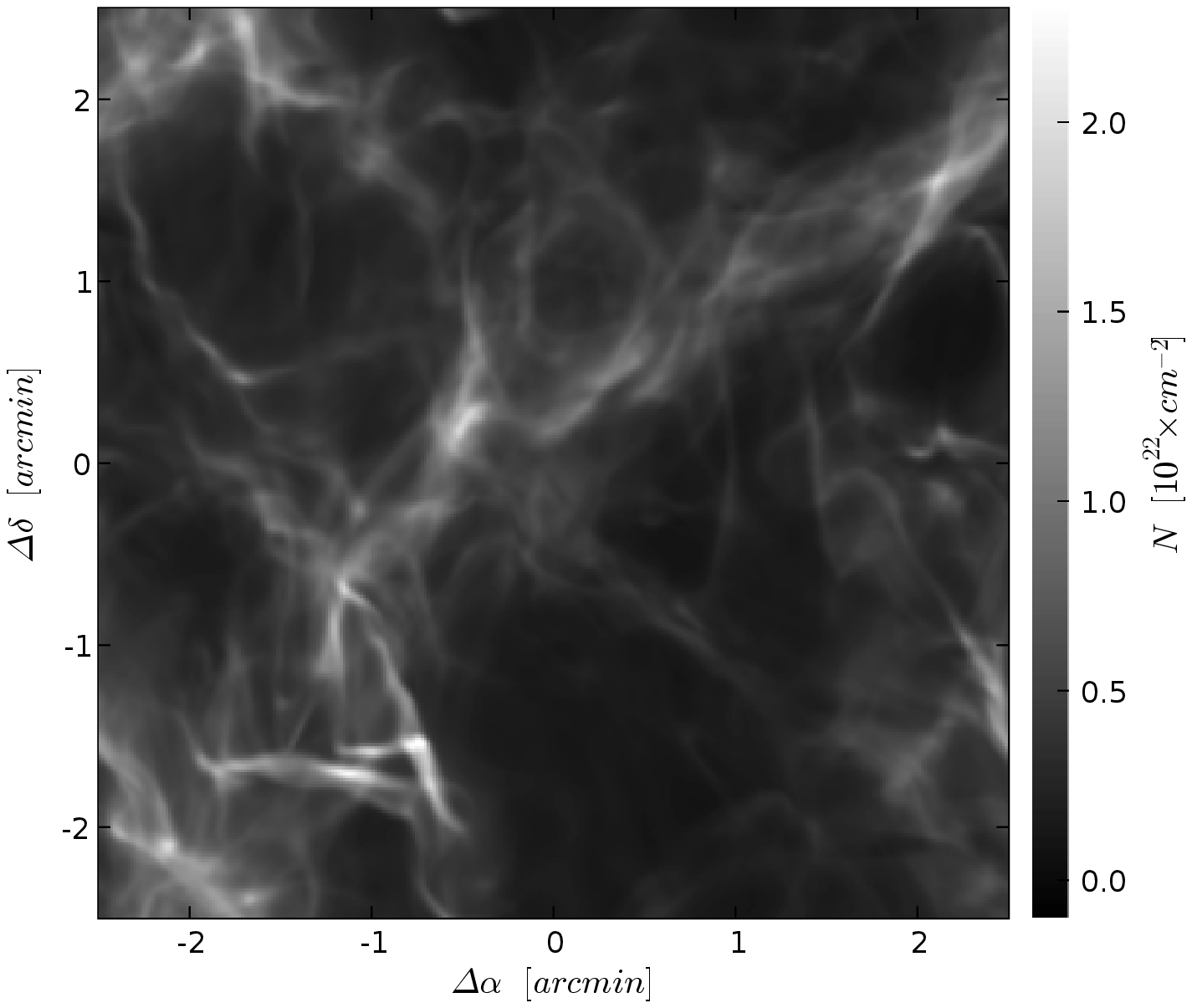}
\caption[]{
{\em Left:}
Ratio of estimated to true column density as a function of true
column density of the model cloud. The dots show estimates based on
scaling linearly the H--band intensity. The plus signs are {\it corrected}
values based on J, H, and K bands.
{\em Right:} The recovered {\it corrected} column density map.
}
\label{fig:NN}
\end{figure*}
Figure~\ref{fig:NN} shows the ratio between the estimated column
densities and the true column density. The uncorrected values obtained 
assuming a linear relation between the H--band intensity and the column 
density are also shown. In regions with $A_{\rm V}\sim 10$~mag the correction 
is almost a factor of three, or a factor $\sim$2 if compared with uncorrected 
K--band intensities. Without correction, the estimated column densities
are on the average approximately 25\% below the correct value. 
Fig.~\ref{fig:NN} shows that the {\it corrected} estimates remain unbiased 
even at large extinctions, $A_{\rm V}\sim 10$~mag. Their mean value deviates 
only 0.2\% from the correct one, and the scatter around the mean is 10\%. 

In Fig.~\ref{fig2}a the scatter is real (no noise was added there) and 
due mostly to spatial variations in the field strength within the cloud. 
The field becomes more attenuated toward the cloud center, reducing the 
ratio of surface brightness to column density. In other words, the 
$a$ coefficients in Eq.~\ref{eq:c1} are functions of position. 
This is the dominant source of the scatter in Fig.~\ref{fig:NN} as well. 
Model calculations may be used to estimate the large scale variations 
in the field strength, and to remove most of this scatter. A detailed 
radiative transfer calculation could also be carried out, where the 
cloud model is iteratively built to replicate the observations as close
as possible. The estimated column densities would then only be subject 
to uncertainties in the cloud structure along the line of sight. Other 
effects could be modeled as well, for example a non--isotropic radiation 
field.

\section{Conclusions}

The average surface brightness of scattered light in the model with mean gas density
$\langle n _{\rm H}\rangle=1000$~cm$^{-3}$ was 
$7.2\times 10^{-19}$, $4.9\times 10^{-19}$ and 
$2.8\times 10^{-19}$~erg\,cm$^{-2}$\,Hz$^{-1}$\,sr$^{-1}$ 
in the J, H and K bands respectively. 
For a target signal--to--noise of 15, 10, and 7 for the J,
H, and K bands respectively, the necessary integration times for the
SOFI instrument on the 3.6\,m NTT telescope become 33, 80, and 250
hours. However, the instrument has 0.29$\arcsec$ pixel size, and if
resolution is degraded to 0.6$\arcsec$ the quoted signal--to--noise
ratios can be reached with less than 100 hours of integration.
With ISAAC/VLT, the same noise level per 0.15$\arcsec$ pixel requires
22+72+196=290 hours. For a resolution decreased to 0.6$\arcsec$, 
the integration time is reduced to $\sim$17 hours.

At a distance of 300~pc, the angular resolution of 0.1$\arcsec$ that 
can be achieved with adaptive optics, corresponds to a spatial resolution
of $\sim30$~AU. Imaging of near--IR scattering is
therefore capable of probing the structure of nearby interstellar
clouds at unprecedented spatial resolution, down to a physical scale
of the order of the size of circumstellar disks, the gyroradius of GeV
cosmic rays or the Kolmogorov turbulence dissipation scale. 

The right panel of Figure~\ref{fig2} shows the H band of our model. 
The image is composed of $250\times 250$ pixels. After decreasing 
the resolution to 0.6$\arcsec$, the same noise level and angular 
dynamical range is reached with ISAAC/VLT in $\sim$17 hours. This 
example shows that imaging of near--IR scattering is a promising method also for
relatively large scale surveys of interstellar clouds, especially with
the new generation of wide--field instruments, such as WFCAM on UKIRT 
and WIRCam on CFHT. As discussed in
the previous section, the method can provide a rather accurate estimate
of the column density, preferentially at values of visual extinction
between approximately 1 and 20 magnitudes, and even below one
magnitude if complemented with shorter wavelengths. It is therefore
ideal for the study of low and medium density interstellar clouds, and
less suitable for the interior of very dense protostellar cores.

\acknowledgements

We are grateful to Jonathan Foster and Alyssa Goodman for pointing our attention
to their beautiful images of scattered light in the Perseus region and for
useful discussions. NASA and NSF grants. P.P. was partially supported by the 
NASA ATP grant NNG056601G and the NSF grant AST-0507768. M.J. and V.-M.P. 
acknowledge the support of the Academy of Finland Grants no. 206049 and 107701.








\end{document}